\documentclass[letterpaper, 10 pt, conference]{ieeeconf}
\usepackage{graphicx}
\usepackage{balance}
\usepackage{comment}
\usepackage{cite}
\usepackage{amssymb}
\usepackage[tight,footnotesize]{subfigure}
\usepackage[active]{srcltx}
\usepackage{amsmath}
\usepackage{amsfonts}

\newcommand{\norm}[1]{\left\lVert#1\right\rVert}
\graphicspath{{./figs/}}
\renewcommand{\baselinestretch}{0.95}
\usepackage{color,soul}
\usepackage{eurosym}
\usepackage{algorithm}
\usepackage{algorithmic}
\usepackage{mathtools}
\IEEEoverridecommandlockouts
\begin{document}
\title{Probabilistic Search and Track with Multiple Mobile Agents}

\author{Savvas~Papaioannou,~Panayiotis~Kolios,~Theocharis~Theocharides,\\~Christos~G.~Panayiotou~ and ~Marios~M.~Polycarpou % <-this % stops a space
\thanks{The authors are with the KIOS Research and Innovation Centre of Excellence, and the Department of Electrical and Computer Engineering, University of Cyprus, {\tt\small \{papaioannou.savvas, pkolios, ttheocharides, christosp, mpolycar\}@ucy.ac.cy}}
}

\maketitle

\begin{abstract}
In this paper we are interested in the task of searching and tracking multiple moving targets in a bounded surveillance area with a group of autonomous mobile agents. More specifically, we assume that targets can appear and disappear at random times inside the surveillance region and that their positions are random and unknown. The agents have a limited sensing range, and due to sensor imperfections they receive noisy measurements from the targets. In this work we utilize the theory of random finite sets (RFS) to capture the uncertainty in the time-varying number of targets and their states and we propose a decision and control framework, in which the mode of operation (i.e. \textit{search} or \textit{track}) as well as the mobility control action for each agent, at each time instance, are determined so that the collective goal of searching and tracking is achieved. Extensive simulation results demonstrate the effectiveness and performance of the proposed solution. 
\end{abstract}

\section{Introduction}
The deployment of multiple autonomous drones or unmanned aerial vehicles (UAVs) in rescue situations is an application scenario that has recently attracted considerable interest \cite{Floreano2015}. Indeed the recent advances in small UAVs and the cost reduction of electronic components have spurred an unprecedented interest in autonomous multi-agent systems for group missions \cite{Rizk2018}. Of particular interest in this work and motivating scenario is the use of a team of autonomous mobile agents in search and rescue missions. More specifically, we believe that a team of autonomous rescue drones can provide an important aid in many search and rescue operations by efficiently searching the area for victims or targets of interest and by accurately tracking the discovered targets until further assistance is available.

 In this work, a probabilistic approach based on multi-object state estimation using random finite sets (RFS) \cite{Mahler2014book} is derived to jointly address the problem of searching and tracking multiple moving targets by a team of autonomous mobile agents. This is a challenging problem since a) the mobile agents exhibit a limited sensing range thus there is a need for efficient searching, b) due to sensor imperfections the agents receive noisy measurements and false alarms which requires efficient algorithms for estimating the number of targets and their states and c) finally the problem becomes even more challenging due to the tradeoff which arises on how many resources to allocate in each task i.e. \textit{searching} or \textit{tracking} at each time instance so that the joint search-and-track objective is achieved as best as possible. The contributions of this work are the following:
\begin{itemize}
	\item Provides a novel probabilistic framework for the problem of joint search-and-track with a team of autonomous mobile agents by accounting the uncertainty in the number of targets, the target dynamics, the target detection process and the limited sensing range for target search.
	\item Develops a novel decision (i.e. the agents can switch between \textit{search} and \textit{track} mode during the mission) and control (i.e. controlling the movement of all agents) algorithm that optimizes the overall search-and-track objective.
	\item Demonstrates the performance of the proposed approach through extensive simulated scenarios.
\end{itemize}

\noindent The rest of the paper is organized as follows. Section \ref{sec:Related_Work} reviews the existing literature on searching and tracking by single and multiple agents, demonstrating in the way the novelty of this work. Section \ref{sec:Problem_Definition} outlines the problem and introduces the proposed system architecture. Section \ref{sec:Background} provides a review on stochastic filtering and random finite sets (RFS) and Section \ref{sec:Problem_Formulation} develops the problem formulation and describes the details of the proposed approach. Finally, Section \ref{sec:Evaluation} conducts an extensive performance analysis and Section \ref{sec:Conclusion} concludes the paper and discusses future work.

\begin{figure*}
	\centering
	\includegraphics[width=\textwidth]{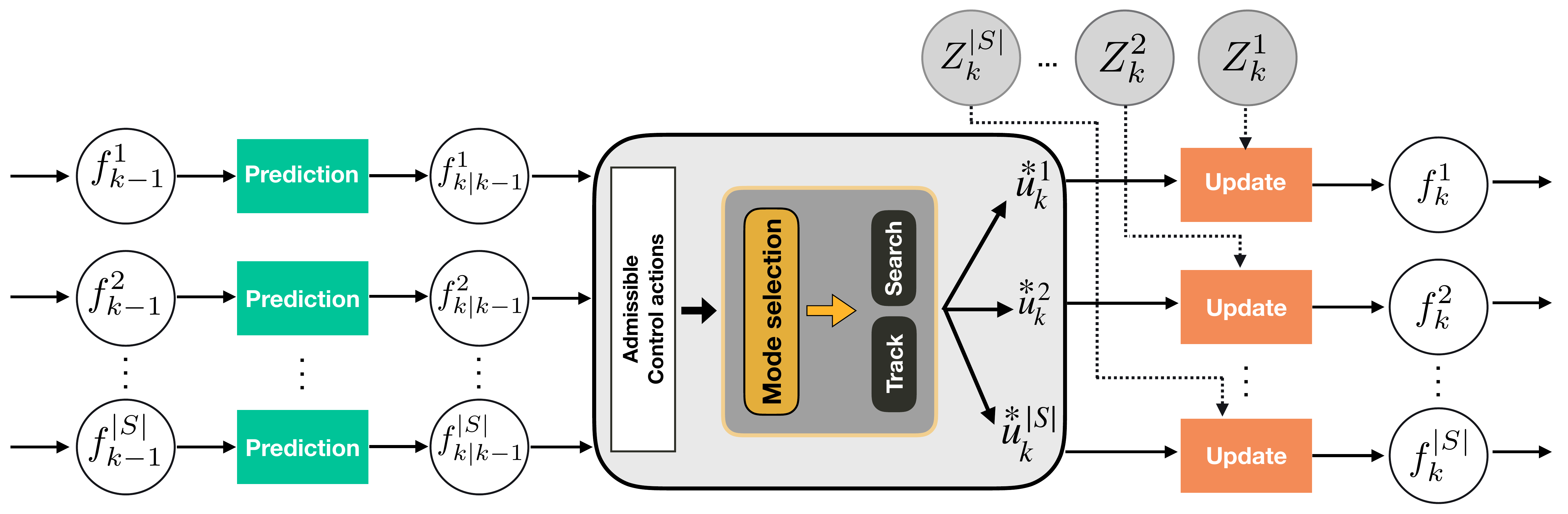}
	\caption{The figure illustrates the proposed system architecture. The proposed approach uses the following recursion: a) for each agent $j$ the predictive multi-object probability distribution $f^j_{k|k-1}$ is first computed from the posterior $f^j_{k-1}$ in the $k_\text{th}$ time-step, b) the number of targets and their states are extracted from each predictive multi-object distribution, c) a decision and control problem is then solved for all agents to decide the mode of operation and control actions, d) finally the agents apply the optimal control actions and move to their new states where they receive target measurements and compute the posterior multi-object density $f^j_{k}$ through the measurement update step.}	
	\label{fig:sys_arch}
\end{figure*}

\section{Related Work}
\label{sec:Related_Work}

Coordinated teams of agents unlock significantly greater capabilities than what is possible by soloing. Taking this fact into consideration, existing literature has looked into several challenging problems that multi-agent systems can successfully tackle including the use of drone swarms in emergency response operations \cite{Kyrkou2019}. The coverage/search problem is among the most cited problems to tackle whereby a fleet of airborne agents need to spread across an area over the shortest time interval for situational awareness or when searching for particular objects. The task assignment, scheduling and path planning problems, considering physical resource constraints such as the total number of agents, their battery levels and their communication ranges have also been considered lately, \cite{Vera2015} \cite{Khan2015} \cite{Avellar2015}. Another challenging problem is that of tracking of single or multiple targets. For this problem, the majority of strategies seek to minimize the tracking error for the detected targets \cite{Li2017} \cite{Dutta2017} \cite{Shahrampour2017}.

%been considered lately, \cite{Vera2015} \cite{Khan2015} \cite{Avellar2015} \cite{Ding2017} \cite{Hayat2016} \cite{Bai2017}. Another challenging problem is that of tracking of single or multiple targets. For this problem, the majority of strategies seek to minimize the tracking error for the detected targets \cite{Li2017} \cite{Dutta2017} \cite{Shahrampour2017} \cite{Xi2016} \cite{Meyer2015}.

In this work our focus in on the joint search-and-track problem which arises especially during and in the aftermath of large-scale disasters. Searching is necessary to detect targets before accurately tracking them, thus efficient multi-agent searching techniques are of high importance. In addition in many real-world search-and-track scenarios the ability to keep track of multiple targets as accurately as possible is essential. Interesting works on the problem of search-and-track are in \cite{Bourgault2006} and \cite{Liu2017}. In \cite{Bourgault2006} the authors develop a probabilistic approach for the search-and-track problem however, they consider only the single-target single-agent scenario. Similarly, a more recent work \cite{Liu2017} investigates the same problem and proposes a model-predictive control (MPC) framework for a single-agent. Initial works on the problem of multi-agent search-and-track are discussed in \cite{Hoffmann2006,Hoffmann2010} but only for stationary targets. The work in \cite{Pitre2012} mainly focuses on path planning algorithms for searching and tracking without considering the problems of false alarms and data association. Finally, a more recent work in \cite{Peterson2017} focuses on the scalability issue when a large number of agents are present and proposes a computationally efficient clustering algorithm. 

Complementary to the aforementioned studies, this work advances the state-of-the-art, by firstly proposing a probabilistic multi-agent framework which accounts for the uncertainties in the number of targets, target dynamics, measurement models and false-alarms and secondly by dynamically adjusting the mode of operation (i.e. search or track) and by determining the control actions of multiple agents so that the collective search-and-track objective is achieved.

\section{Problem Definition and System Overview}
\label{sec:Problem_Definition}

In this work we assume that a bounded region of interest needs to be searched for potential targets by a group of mobile agents. The targets can enter and exit the surveillance region at random times (i.e. the number of targets is not fixed and needs to be estimated) and their positions are unknown. In addition, the agents are equipped with imperfect sensors and thus they detect targets within their limited sensing range with probability less than one. Moreover, the agents receive noisy measurements from the targets and at random times they receive spurious measurements (e.g. false alarms/clutter) as well. 
The resulting problem we would like to address is the following: 

\textit{At every discrete time-step, a team of mobile agents must jointly decide their mode of operation (i.e. search or track) and select their mobility control actions in order to find the best tradeoff between a) maximizing the area coverage and b) accurately tracking all detected targets}.

The proposed approach is based on the framework of multi-object (or multi-target) filtering with random finite sets \cite{Mahler2014book}. This framework extends the stochastic filtering theory \cite{Chen2003} to allow us to jointly estimate both the number and the states of an unknown number of objects (or targets) from a set of measurements in the presence of clutter. In this section we provide an overview of the proposed system architecture, as shown in Fig. (\ref{fig:sys_arch}).

Let $S = \{1,2,...,|S|\}$ be the set of available mobile agents. Each agent $j$ can operate in either \textit{search} or \textit{track} mode and in each mode the agent is subject to a set of admissible control actions, denoted by $\mathbb{U}^j_{k}$ for time $k$. When a subset of agents $\alpha \subseteq S$ is in \textit{search} mode the collective objective over these agents is the maximization of the covered area. In other words the agents should traverse the surveillance area in such a way so that the coverage is maximized. On the other hand, when a subset of agents $\alpha^\text{C} \subseteq S$ (where $\alpha^\text{C}$ denotes the complement of $a$ with respect to $S$) is in \textit{track} mode, the objective is the maximization of the tracking performance, i.e. the number of targets and their states inside the surveillance area should be estimated as accurately as possible.
The cost of having a subset of agents in \textit{search} or \textit{track} mode is given by the objective functions $\xi_\text{search}(\alpha)$ and $\xi_\text{track}(\alpha^\text{C})$, respectively. The proposed controller solves the following:

  \emph{At each time-step $k$ it finds the subsets of agents $\alpha \subseteq S$ and $\alpha^\text{C} \subseteq S$ for searching and tracking, respectively, and the appropriate control actions $(u^j_k \in \mathbb{U}_k^j ~ \forall j)$ for each agent so that the total cost of searching and tracking $\xi_\text{search}(\alpha) +\xi_\text{track}(\alpha^\text{C})$ is minimized.}

In order to achieve this, the proposed system takes the following steps: Each agent maintains a multi-object probability distribution which jointly accounts for the uncertainty in the number of targets and their states (i.e. position and velocity) inside its sensing range (this multi-object probability distribution is a generalization of the probability density function on random finite sets). Let this posterior distribution at time $k-1$ be denoted by $f^j_{k-1}$ for agent $j$. In the first step, the predictive density $f^j_{k|k-1}$ is obtained through a prediction step, as shown in Fig. \ref{fig:sys_arch}. Then the predicted number of targets and their states are extracted from the predictive density and based on these predictions, the proposed controller finds the combination of hypothesized control actions $(u^j_k \in \mathbb{U}_k^j ~ \forall j)$ as well as the mode of operation of each agent which minimizes the total cost of the search-and-track objective. The optimal hypothesized control actions $u^{*1}_\text{k},...,u^{*|S|}_\text{k}$ are then applied to the agents. The agents move to their new states at time $k$ where they can receive the measurements sets $Z^1_k,...,Z^{|S|}_k$ from the targets (if any). These measurement sets at time $k$ are then used in the measurement update step i.e., each agent uses the received measurements to compute the posterior multi-object distribution $f^j_k$ from which the posterior number of targets and their states can be estimated. 

The same procedure is repeated recursively over time. We should note here that the number of targets (and their states) is being estimated recursively over time and that based on this estimate the mode of operation should be carefully chosen. In the next section, we provide a brief overview on single and multi-object stochastic filtering which will help us set the basis of the proposed approach.

\section{Background on Stochastic Filtering}
\label{sec:Background}
In this section we provide a brief overview on the theory of random finite sets (RFS) and stochastic filtering that will be used this framework. A more detailed description can be found in \cite{Simon2006,Mahler2014book}.

\subsection{Stochastic Filtering}
In stochastic filtering, we are interested in the posterior probability density of the hidden state $x_k$ at time $k$ given all measurements $z_{1:k}=(z_1,...,z_k)$ up to time $k$, denoted by:

\begin{equation}
	p_k(x_k|z_{1:k})
\end{equation}

\noindent In this paper, variable $x_k$ denotes the state of a single target at time $k$. Given an initial density on the target state $p_0(x_0)$, the posterior density at time $k$ can be computed using the Bayes recursion \cite{Simon2006} as:
\begin{align} 
p_{k|k-1}&(x_k|z_{1:k-1})= \label{eq:predict}\\
& \int  \pi_{k|k-1}(x_k|x_{k-1}) ~ p_{k-1}(x_{k-1}|z_{1:k-1}) ~d x_{k-1} \notag
\end{align}
%\vspace{-6mm}
\begin{align}
p_k(x_k|z_{1:k}) = \frac{g_k(z_k|x_k) ~ p_{k|k-1}(x_k|z_{1:k-1})}{\int g_k(z_k|x_k) ~ p_{k|k-1}(x_k|z_{1:k-1}) ~dx_k} \label{eq:update}
\end{align}

\noindent where Eqn. (\ref{eq:predict}) and (\ref{eq:update}) are referred to as the prediction and measurement update steps, respectively. In the prediction step, the Chapman-Kolmogorov equation is used to calculate the predictive density $p_{k|k-1}(x_k|z_{1:k-1})$ i.e. the predicted probability density function (pdf) of $x_k$ based on the measurements up to time $k-1$. The measurement update step applies the Bayes rule to update the predicted pdf using the current measurement $z_k$ and obtain the posterior distribution $p_k(x_k|z_{1:k})$ for the state $x_k$. The function $\pi_{k|k-1}(x_k|x_{k-1})$ models the uncertainty of the current state given the previous state and is referred to as the transitional density and the function $g_k(z_k|x_k)$ describes the distribution of measurements given the current state and is referred to as the measurement likelihood function. Finally, once the posterior density is obtained, the state $x_k$ can be extracted using for instance the expected a posteriori (EAP) or the maximum a posteriori (MAP) estimators.

\subsection{Random Finite Sets}
A random finite set \cite{Mahler2014book,Mahler2013,Mahler2004_2} (RFS) is a finite-set-valued random variable, which differs from a random vector in two ways: a) the number of elements in a RFS is random and b) the order of the elements in a RFS is irrelevant. More specifically, a RFS $X=\{x_1,x_2,...,x_n\}$ is completely specified by a) its cardinality distribution $\rho(n) = p(|X|=n)$ which defines the probability over the number of elements in $X$ and b) by a family of conditional joint probability distributions $p(x_1,...,x_n|n)$, $x_1,...,x_n \in \mathcal{X}$ over the distribution of its elements. The belief density or otherwise known multi-object/multi-target pdf $f(X)$ of the RFS $X$ is given by: $f(X) = f(\{x_1,...,x_n\}) = n!  \rho(n) p(x_1,...,x_n|n)$ and the notion of integration is given by the set-integral which is defined as:
\begin{equation}
	\int f(X) \delta X = f(\emptyset)+\sum^{\infty}_{n=1} \frac{1}{n!} \int f(\{x_1,...,x_n\})d_{x_1}...d_{x_n}
\end{equation}

\noindent That said, it has been shown \cite{Mahler2007book} that the stochastic filtering recursion of Eqn. (\ref{eq:predict})-(\ref{eq:update}) can be extended for RFSs in order to allow for the recursive computation of the multi-object posterior density $f_k(X_k|Z_{1:k})$ of RFS $X_k$ given RFS measurements $Z_{1:k}$. In this recursion the random variables $x_k$ and $z_k$ are represented by the RFSs' $X_k$ and $Z_k$, the functions $\pi_{k|k-1}(x_k|x_{k-1})$ and $g_k(z_k|x_k)$ are replaced by the multi-object transition density and the multi-object likelihood function respectively and finally the integrals have become set integrals as defined above. This recursion (i.e. the multi-object stochastic filtering) however, has in general no analytic solution, and because of its combinatorial complexity is usually intractable. Thus, approximations exist which propagate only summary statistics, instead of the full posterior density.

A prominent approach which is summarized here is the multi-Bernoulli filter \cite{Vo2009}, which is based on the approximation of the multi-object posterior density $f_k(X_k|Z_{1:k})$ as multi-Bernoulli pdf. In essence this approach only propagates the multi-Bernoulli parameters $\{(r^i_{k},p^i_{k}(x))\}^{v_{k}}_{i=1}$ in time, using a predictor-corrector scheme with computational complexity $O(mn)$, where $m$ is the number of measurements and $n$ is the number of targets.

 More specifically, let the multi-object posterior density at time $k$ be approximated by a multi-Bernoulli distribution $f_{k}= \{(r^i_{k},p^i_{k}(x))\}^{v_{k}}_{i=1}$ where $v_k$ is the maximum number of hypothesized targets at time $k$, $r^i_k$ denotes the probability that the $i_\text{th}$ target actually exists and $p^i_{k}(x)$ is the spatial density of the $i_\text{th}$ target. The predictive multi-object density at time $k$ is given by: $f_{k|k-1} = f^\text{persist}_{k|k-1} \cup f^\text{birth}_{k}$ where $f^\text{persist}_{k|k-1} = \{(r^i_{k|k-1},p^i_{k|k-1}(x))\}^{v_{k-1}}_{i=1}$ describes the set of persisting targets from the previous time-step and $f^\text{birth}_{k} = \{(r^i_{k},p^i_{k}(.))\}^{\Gamma_{k}}_{i=1}$ are the parameters of a multi-Bernoulli RFS for the new-born targets at time $k$. The distribution of persisting targets $f^\text{persist}_{k|k-1}$ is given by:
\begin{align}
r^i_{k|k-1} &= r^i_{k-1}\left< p^i_{k-1} , p_{S}\right>  \label{eq:RFS_prediction2}\\
p^i_{k|k-1}(x) &= \frac{\left< \pi_{k|k-1}(x) , p^i_{k-1}  p_{S} \right>}{\left< p^i_{k-1} , p_{S}\right>} \label{eq:RFS_prediction3}
\end{align}
where the notation $<f,g>$ is defined as $\int f(x)g(x)dx$ and $p_S(x)$ denotes the probability of target survival.  
Given the predictive multi-object density $f_{k|k-1} = \{(r^i_{k|k-1},p^i_{k|k-1}(x))\}^{v_{k|k-1}}_{i=1}$ (where the term $v_{k|k-1}$ denotes the number of hypothesized Bernoulli components after accounting for the target births) and the new measurement set $Z_k=\{z^1_k,...,z^{m_k}_k\}$ then the posterior multi-object distribution at time $k$ is approximated by the multi-Bernoulli distribution $f_k = f^\text{legacy}_{k} \cup f^\text{meas}_{k}$ where $f^\text{legacy}_{k} = \{(r^i_{k},p^i_{k}(x))\}^{v_{k|k-1}}_{i=1}$ contains legacy targets i.e. updates of predicted targets, assuming that no measurements were collected from them and given by \cite{Vo2009}:
\begin{align}
r^i_{k} &= r^i_{k|k-1} \frac{1 - \left< p^i_{k|k-1} , p_{D}\right>}{1 - r^i_{k|k-1}\left< p^i_{k|k-1} , p_{D}\right>} \label{eq:r_update_legacy} \\
p^i_{k}(x) &= p^i_{k|k-1}(x) \frac{1-p_{D}(x)}{1 - \left< p^i_{k|k-1} , p_{D}\right>} \label{eq:p_update_legacy}
\end{align}

\noindent where $p_{D}(x)$ is the probability of detecting a target with state $x$. $f^\text{meas}_{k} = \{(r_{k}(z),p_{k}(x,z))\}_{z \in Z_k}$ contains new targets i.e. joint updates of all predicted targets using each of the measurements separately and given by:
\begin{align}
r_{k}(z) = \frac{\sum \displaylimits^{v_{k|k-1}}_{i=1} \frac{r^i_{k|k-1}(1-r^i_{k|k-1})\left< p^i_{k|k-1} , L^z_k\right> }{\left (1-r^i_{k|k-1} \left<p^i_{k|k-1} , p_{D}\right> \right)^2}}{\kappa_k(z) + \sum \displaylimits^{v_{k|k-1}}_{i=1} \frac{r^i_{k|k-1}\left< p^i_{k|k-1} , L^z_k\right>}{1-r^i_{k|k-1} \left<p^i_{k|k-1} , p_{D}\right>}} \label{eq:r_update_meas}\\
p_{k}(x,z) = \frac{\sum \displaylimits^{v_{k|k-1}}_{i=1} \frac{r^i_{k|k-1}}{1-r^i_{k|k-1}} p^i_{k|k-1} L^z_k(x)}{\sum \displaylimits^{v_{k|k-1}}_{i=1} \frac{r^i_{k|k-1}}{1-r^i_{k|k-1}} \left< p^i_{k|k-1} \cdot L^z_k\right> } \label{eq:p_update_meas}
\end{align}
where $L^z_k(x) = g_k(z|x) p_{D}(x)$.

\section{System Model}
\label{sec:Problem_Formulation}
In this section we discuss the details of the proposed approach. More specifically in this paper we make the following modeling assumptions:

\subsection{Single Target Dynamics}
Lets assume first that the target's state vector $x_k \in \mathcal{X}$ evolves in time according to the following equation:
\begin{equation}\label{eq:single_dynamics}
	x_k = \zeta_{k}(x_{k-1}) + \text{v}_{k}
\end{equation} 

\noindent where the possibly non-linear function $\zeta_{k}$ which is usually referred to as the dynamical model, describes the evolution of the state vector as a first order Markov process with transitional density $\pi(x_k|x_{k-1}) = p_\text{v}(x_k - \zeta_{k}(x_{k-1}) )$. The random process $\text{v}_{k} \in \mathbb{R}^{n_x}$, which is referred to as the process noise, is independent and identically distributed (IID) according to the probability density function $p_\text{v}$. The role of the process noise is to model random disturbances in the evolution of the state. In this paper the state vector $x_k$ is composed of position and velocity components i.e. $x_k = [\text{x},\dot{\text{x}},\text{y},\dot{\text{y}}]^\top$ in Cartesian coordinates.

\subsection{Agent dynamics}
Let $S = \{1,2,...,|S|\}$ be the set of all mobile agents that operate in a discrete-time setting.
At time $k$, the 2D surveillance area $\mathcal{A} \subset \mathbb{R}^2$ is monitored by $|S|$ mobile agents with states $s^1_k,s^2_k,...,s^{|S|}_k$ taking values in $\mathcal{A}$. Each agent is subject to the following dynamics with bounded control inputs:
\begin{equation} \label{eq:controlVectors}
s^j_{k} = s^j_{k-1} + \begin{bmatrix}
						l_1\Delta_R \text{cos}(l_2 \Delta_\theta)\\
						l_1\Delta_R \text{sin}(l_2 \Delta_\theta)
					\end{bmatrix},  
					\begin{array}{l} 
						l_2 = 0,...,N_\theta\\ 
						l_1 = 0,...,N_R
				    \end{array} 
\end{equation}
where  $s^j_{k-1} = [s^j_x,s^j_y]^\top_{k-1}$ denotes the position (i.e. xy-coordinates) of the $j_{\text{th}}$ agent at time $k-1$, $\Delta_R$ is the radial step size, $\Delta_\theta=2\pi/N_\theta$ and the parameters $(N_\theta,N_R)$ control the number of possible control actions. We denote the set of all admissible control vectors (or actions) of agent $j$ at time $k$ with $\mathbb{U}^j_{k}=\{s^{j,1}_{k},s^{j,2}_{k},...,s^{j,|\mathbb{U}_{k}|}_{k}  \}$ as computed by Eqn. (\ref{eq:controlVectors}). 

%\begin{figure}
%	\centering
%	\includegraphics[scale=0.65]{fcontrols3d.pdf}
%	\caption{The figure shows an illustrative example of the agent's admissible control actions $\mathbb{U}_k$ at time $k$. In this example the radial displacement $\Delta_R=5$m, $N_R=2$ and $N_\theta = 8$ which gives a total of 17 control actions, including the initial position of the agent at time $k-1$ which is marked with a circle.}
%	\label{fig:controls}
%\end{figure}

\subsection{Agent sensing model}
The ability of an agent to sense its 2D environment is modeled by the function $p_D(x_k,s_k)$ that measures the probability that a target with state $x_k$ at time $k$ is detected by an agent with state (i.e. 2D position coordinates) $s_k =[s_x,s_y]^\top_k \in \mathcal{A} \subseteq R^2$. More specifically the detection probability of a target with state $x_k$ at position $\textit{p}_k = Hx_k$ (where $H$ is a matrix that extracts the $xy$-coordinates of a target from its state vector) from an agent with state $s_k$, is given by:
\begin{equation}\label{eq:sensing_model}
 p_D(x_k,s_k) = 
  \begin{cases} 
   p^{\max}_D & \text{if } d_k < R_0 \\
   \text{max}\{0, p^{\max}_D - \eta(d_k-R_0) \} & \text{if } d_k \ge  R_0
  \end{cases}
\end{equation}
where $d_k=\norm{H x_k-s_k}_2$ denotes the Euclidean distance between the agent and the target, $p^{\max}_D$ is the detection probability for targets that reside within $R_0$ distance from the agent's position and finally $\eta$ captures the reduced effectiveness of the agent to detect distant targets. In addition, when the agent detects a target, it receives a measurement vector $z_k \in \mathcal{Z}$ (composed of range and bearing measurements) which is related to the target state as follows:
\begin{equation}\label{eq:meas_model}
	z_k = h_{k}(x_k,s_k) + w_k
\end{equation}
where the non-linear function $h_{k}$ projects the state vector to the observation space. The random process $\text{w}_{k} \in \mathbb{R}^{n_z}$ is IID, independent of $\text{v}_{k}$ and distributed according to $p_\text{w}$. The probability density of receiving measurement $z_k$ for a target with state $x_k$ is given by the measurement likelihood function $g(z_k|x_k,s_k) = p_\text{w}(z_k - h_k(x_k,s_k))$.

\subsection{Multi-Target Dynamics and Sensor Observation Models}
Suppose now that multiple targets $x_k^i$ exists inside the surveillance region at a particular time instance $k$ where we assume that the targets are independent of each other. The targets can spawn from anywhere in the surveillance area and target births and deaths occur at random times. This means that at each time $k$, there exist $n_k$ target states $x^1_k, x^2_k,...,x^{n_k}_k$, each taking values in the state space $\mathcal{X}$ where both the number of targets $n_k$ and their individual states $x_k^i, \forall i$ are random and time-varying. The group of target states at time $k$ can now be modeled as the RFS $X_k = \{x^1_k, x^2_k,...,x^{n_k}_k\} ~ \in \mathcal{F}(\mathcal{X})$ where $\mathcal{F}(\mathcal{X})$ denotes the space of all finite subsets of $\mathcal{X}$. Note here that $n_k$ is the true but unknown number of targets at time $k$ which needs to be estimated along with the target states $x_k^i, \forall i$.

Let $x_{k-1} \in X_{k-1}$ denote the state vector of a single target, and $X_{k-1}$ the multi-target state (i.e. RFS) at time $k-1$. The multi-target dynamics is given by:
\begin{equation}\label{eq:RFS_transition_model}
	X_k = \left[ \underset{x_{k-1} \in X_{k-1}}{\bigcup} \Psi(x_{k-1}) \right] \cup \Gamma_k
\end{equation}
where $\Psi(x_{k-1})$ is a Bernoulli RFS \cite{Ristic2013} which models the evolution of the set from the previous state, with parameters $r = p_{S,k}(x_{k-1})$ and $p(x_k) = \pi(x_k|x_{k-1})$. Thus a target with state $x_{k-1}$ continues to exist at time $k$ with surviving probability $p_{S,k}(x_{k-1})$ and moves to a new state $x_k$ with transition probability $\pi(x_k|x_{k-1})$. Otherwise the target dies with probability $1-p_{S,k}(x_{k-1})$. The term $\Gamma_k$ denotes the multi-Bernoulli RFS of spontaneous births \cite{Vo2009}. 

An agent can receive $m_k$ measurements $z^1_k, z^2_k,...,z^{m_k}_k$, from the targets and clutter, each taking values in the measurement space $\mathcal{Z}$, where $m_k$ and $z^i_k, \forall ~i$ are random and time varying. Thus the group of received measurements by an agent at time $k$ can be modeled as the RFS $Z_k = \{z^1_k, z^2_k,...,z^{m_k}_k\} ~ \in \mathcal{F}(\mathcal{Z})$ where $\mathcal{F}(\mathcal{Z})$ denotes the space of all finite subsets of $\mathcal{Z}$. As a consequence the agent $j$ with state $s_k^j \in \mathcal{A}$ receives measurements at time $k$,  $z_k^{ji} \in Z_k^j$ where $Z_k^j$ is the RFS multi-target measurement set described by:
\begin{equation}\label{eq:RFS_measurement_model}
	Z_k^j = \left[ \underset{x_{k}^j \in X_{k}^j}{\bigcup} \Theta(x_{k}^j) \right] \cup \text{K}_k
\end{equation}
where $\Theta(x_{k}^j)$ is a Bernoulli RFS which models the target generated measurements with parameters $(p_{D}(x^j_k,s^j_k),g_k(z^{ji}_k|x^j_k,s^j_k))$. Thus a target with state $x^j_k$ at time $k$ is detected by the $j_\text{th}$ agent with state $s^j_k$ with probability $p_{D}(x^j_k,s^j_k)$ (see Eqn. (\ref{eq:sensing_model})) and generates a measurement $z^{ji}_k$ with likelihood $g_k(z^{ji}_k|x^j_k,s^j_k)$ (see Eqn. (\ref{eq:meas_model})) or is missed with probability $1-p_{D}(x^j_k,s^j_k)$ and generates no measurements. The term $\text{K}_k$ is a Poisson RFS \cite{Ristic2013} which models the set of false alarms or clutter received by the sensor at time $k$ with intensity function $\kappa_k(z^{ji}_k) = \lambda f_{c}(z^{ji}_k)$, where in this paper $f_c(.)$ denotes the uniform distribution over $\mathcal{Z}$.

%The transition model in Eqn. (\ref{eq:RFS_transition_model}) jointly incorporates motion, birth and death for multiple targets, while the sensor observation model in Eqn. (\ref{eq:RFS_measurement_model}) jointly accounts for detection uncertainty and clutter. It is also worth noting that $X_k$ is a multi-Bernoulli RFS given $X_{k-1}$ and that the RFSs constituting the union in Eqn. (\ref{eq:RFS_transition_model}) are mutually independent. Similarly, conditional on $X_k$, the target generated measurements in Eqn. (\ref{eq:RFS_measurement_model}) form a multi-Bernoulli RFS. As a consequence, the posterior probability distribution of the multi-target state $X_k$ at time $k$ given all measurement sets $Z_{1:k}$ up to time $k$ can be obtained via the multi-Bernoulli filter expressed by Eqn. (\ref{eq:RFS_prediction1})-(\ref{eq:p_update_meas}).

%Note here that Eqn. (\ref{eq:RFS_measurement_model}) accounts for the detection uncertainty and clutter and that in this work the following holds by design $X_k = \bigcup_j X^j_k, \forall j$ and $X^{j_1}_k \cap X^{j_2}_k = \emptyset, ~ \forall j_1, j_2$ i.e. a target cannot be tracked by more than one agent. 

\section{Multi-Agent Decision and Control}
To allow the agents to switch modes (i.e. \textit{search} or \textit{track}) dynamically during the mission and to optimize the joint search-and-track objective the following optimization problem is to be solved at each time-step $k$:
\begin{equation} 
\label{eq:o1}
(\text{P}_1) ~ \underset{u^j_k, \forall j \in S}{\arg\min} \left[ \xi^\alpha_\text{search}(\{u^j_k : j \in \alpha\}) + \underset{\substack{j=1 \\ j \not\in \alpha }}{\sum^{|S|}} \xi^j_\text{track}(u^j_{k},Z^j_{k|k-1,u^j_{k}})  \right]
\end{equation}
\begin{equation*}
\begin{aligned}
& \text{subject to} & & \alpha \in \mathbb{P}(S) \\
&&& u^j_{k} \in \mathbb{U}^j_{k} ~,\>\forall j \in S\\
&&& Z^j_{k|k-1,u^j_{k}} = Z_\text{PIMS}(\hat{X}^j_{k|k-1},u^j_{k})~,\>\forall j \in S\\
&&& \norm{u_k^i - u_k^j}> d_\text{min}~, \forall ~i \ne j
\end{aligned}
\end{equation*}
\noindent where $\alpha$ is the decision variable for the agents in search mode and $\mathbb{P}(S) =\Big \{ \emptyset,\{1\},\{2\},\dots,\{|S|\},\{1,2\},\dots \Big \}$ denotes the power set of $S$. The function $\xi^\alpha_\text{search}(\{u^j_k : j \in \alpha\})$ takes as input a set of control actions for the agents in search mode and returns the total cost of \text{searching} and the function $\xi^j_\text{track}(u^j_{k},Z^j_{k|k-1,u^j_{k}})$ returns the cost of tracking when the control $u^j_k$ is taken and the predicted multi-target measurement set $Z^j_{k|k-1,u^j_{k}}$ is supposed to have been received by agent $j$. $\hat{X}^j_{k|k-1}$ denotes the predicted multi-target state for time $k$ from agent $j$ based on time $k-1$. Note that the cost of tracking depends on 
the received measurements which in turn depend on the control $u^j_k$ that is applied to agent $j$ (i.e. depending on the control action taken, different measurements will be received). The measurement generator function $Z_\text{PIMS}(\hat{X}^j_{k|k-1},u^j_{k})$ generates the predicted ideal measurement set based on $\hat{X}^j_{k|k-1}$ and $u^j_k$. Finally, the constraint $\norm{u_k^i - u_k^j}> d_\text{min}, \forall ~i \ne j$ does not allow any two agents to get closer than $d_{min}$ from each other. This is by design since in this work there is no benefit of two agents to search the same area or perform tracking of the same target. We should note here that the optimization of Eqn. (\ref{eq:o1}) partitions the agents into two groups i.e. search or track and finds the joint control actions so that the joint search-and-track cost is minimum. Finally, each agent $j$ uses the multi-Bernoulli recursion Eqn. (\ref{eq:RFS_prediction3}) - (\ref{eq:p_update_meas}) to compute posterior multi-object density at time $k$ given by $f^j_k(X_k|Z_{1:k}) = \{(r^{ji}_{k},p^{ji}_{k}(.))\}^{v^j_k}_{i=1}$ from where the estimated number of targets $\hat{n}^j_k$ and the multi-target state $\hat{X}^j_k$ can are computed (see \ref{ssec:track_obj}). The total estimated number of targets at time $k$ in the surveillance area is then computed as $\hat{n}_k = \sum_j \hat{n}^j_k$ and the multi-target state which accounts for all targets in the area as $\hat{X}_k = \bigcup_j \hat{X}^j_k$.

\subsection{Multi-Agent Tracking Objective} \label{ssec:track_obj}

The tracking objective derived hereafter allows the proposed system to determine which controls to assign to each agent in order to optimize the tracking performance. Lets assume that at time $k$ the multi-object target distribution estimated by agent $j$ is approximated by a multi-Bernoulli pdf $f^j_{k} = \{(r^{ji}_{k}, p^{ji}_{k}(x))\}^{v^j_{k}}_{i=1}$. From this distribution the expected a posteriori (EAP) estimate of number of targets $\hat{n}^j_k$ can be computed as 
$\hat{n}^j_k = round(\sum_i r^{ji}_k)$ and the multi-target state $\hat{X}^j_k$ can then be computed by 
taking the $\hat{n}^j_k$ components with the highest probabilities of existence and extracting the single target 
states (i.e. $\hat{x}_k$) from their individual posterior 
densities as $\hat{X}^j_k = \bigcup_{i=1}^{\hat{n}^j_k} \{\hat{x}_k~=~\arg\max~p^{ji}_k(x_k)\}$.

Observe that the sensor measurements $Z_k^j$ play a crucial role in the measurement update step i.e. Eqn. (\ref{eq:r_update_meas})-(\ref{eq:p_update_meas}) for the computation of the posterior distribution. In addition, the \textit{quality} of the received measurements depend on the state $s_k^j$ (i.e. the control action taken) of the agent through the functions $p_{D}(x^j_k,s^j_k)$ and $g_k(z^{ji}_k|x^j_k,s^j_k)$. Since the state $s_k^j$ of an agent is assumed to be controllable via Eqn. (\ref{eq:controlVectors}), we come to the following conclusion: the control action affects the received measurements which in turn affects the estimate of the target state. Thus in order to increase the tracking accuracy it suffices to find an optimized control decision. 

Let us denote the objective function we wish to minimize by $\xi^j_\text{track}(u^j_{k},Z^j_{k}),~ u^j_{k} \in \mathbb{U}^j_{k}$ if we were going to apply at time $k$ the control action $u^j_{k}$ and observe the measurement set $Z_{k}^j$. Notice that the objective function we wish to minimize depends on an unknown future measurement set, i.e. the measurement set will be receive at time $k$ only after applying the control $u^j_{k}$. A common approach around this problem (as discussed in \cite{Clark2011,Beard2015}) is to take the statistical expectation of the objective function with respect to all possible values of future measurements i.e. $\mathbb{E} [\xi^j_\text{track}(u^j_{k},Z^j_{k})]$ thus the optimization problem becomes $u_{k}^{j\star} = \arg\min~ \mathbb{E} \left[ \xi^j_\text{track}(u^j_{k},Z^j_{k}) \right]$. Computing this expectation however, is computationally intensive because it requires the generation of an ensemble  of measurement sets (pseudo-measurements) $Z^j_{k}$ for each hypothesized control action. An alternative and computationally cheaper approach which we adopt in the paper uses the predicted ideal measurement set (PIMS)\cite{Mahler2004} i.e. the noise-free clutter-free measurement set which is most likely to be obtained when a particular control action is applied, denoted $Z_{k|k-1}$ in this paper. Thus the control problem now becomes:
\begin{equation} \label{eq:optimize_uk}
u_{k}^{j\star} = \underset{u^j_{k} \in \mathbb{U}^j_{k}}{\arg\min}~ \xi^j_\text{track}(u^j_{k},Z^j_{k|k-1})
\end{equation}
\noindent The above optimization can now be solved as follows: Using the posterior distribution $f^j_{k-1}$ at time $k-1$ we first compute the predictive density  $f^j_{k|k-1}=\{(r^{ji}_{k|k-1}(u^j_{k-1}),p^{ji}_{k|k-1}(.,u^j_{k-1}))\}^{v^j_{k|k-1}}_{i=1}$ which is obtained without yet performing any control actions. The above notation is to explicitly show the dependence on the control action that was taken for the time-step $k-1$. Next we perform a pre-estimation step where the number of targets $\hat{n}^j_{k|k-1}$ is first estimated from the predictive density by counting the Bernoulli components which exhibit a probability of existence $r^{ji}_{k|k-1}>0.5, \forall i \in \{1...v^j_{k|k-1}\}$. The states of those targets are then extracted from their spatial distributions to obtain $\hat{X}^j_{k|k-1}$. Then for a given hypothesized control action $u^j_{k} \in \mathbb{U}^j_{k}$ the corresponding PIMS $Z^j_{k|k-1  ,u^j_k}$ is generated as:
\begin{align} \label{eq:GenPIMS}
     Z^j_{k|k-1,u^j_k} &= Z^j_{k|k-1,u^j_k} ~~\cup \\
      &\{ \underset{z}{\arg\max}~ g_k(z|\hat{x}^j_{k|k-1},u^j_k)\}, ~ \forall \hat{x}^j_{k|k-1} \in \hat{X}^j_{k|k-1} \notag
\end{align}

\noindent Now for each pair $(u^j_k,Z^j_{k|k-1,u^j_k})$ a pseudo-update is performed using Eqn. (\ref{eq:r_update_legacy}) - (\ref{eq:p_update_meas}) to compute the (pseudo) posterior density $\hat{f}^j_k(X^j_k|Z^j_{k|k-1,u^j_{k}},u^j_{k})$ from where the number of targets is estimated. We consider the variance of this estimate as a measure of uncertainty which should be minimized. More specifically let $\hat{f}^j_k = \{(r^{ji}_{k}(u^j_{k},Z^j_{k|k-1,u^j_{k}}), p^{ji}_{k}(.,u^j_{k},Z^j_{k|k-1,u^j_{k}}))\}^{v^j_{k,u^j_k}}_{i=1}$ then the expected number of targets $\hat{n}^j_k$ and its variance $\sigma^j_k$ is given by:

\begin{align}
	\hat{n}^j_k(u^j_{k},Z^j_{k|k-1,u^j_{k}})  &= \sum_{i=1}^{v^j_{k,u^j_k}} r^{ji}_{k}(u^j_{k},Z^j_{k|k-1,u^j_{k}}) \\
	\sigma^j_k(u^j_{k},Z^j_{k|k-1,u^j_{k}}) &= \sum_{i=1}^{v^j_{k,u^j_k}} r^{ji}_{k}(u^j_{k},Z^j_{k|k-1,u^j_{k}})~ \times \\
		& \quad( 1 - r^{ji}_{k}(u^j_{k},Z^j_{k|k-1,u^j_{k}}) ) \notag 
\end{align}

\noindent In essence, by minimizing the cardinality variance $\sigma^j_k$, the agent $j$ chooses the control action which results in the best estimate regarding the number of targets. This variance is maximized when $r^{ji}_{k}(u^j_{k},Z^j_{k|k-1,u^j_{k}}) = 0.5,\forall i$ thus we can define:

\begin{equation} \label{eq:single_track_objective}
\xi^j_\text{track}(u^j_{k},Z^j_{k|k-1,u^j_{k}}) = 
\begin{dcases}
      \frac{4\sigma^j_k}{v^{j}_{k}} & \text{,~if~} \hat{n}^j_k(u^j_{k},Z^j_{k|k-1,u^j_{k}}) \ge  1 \\
      ~1 & \text{,~otherwise}
   \end{dcases}
\end{equation}   
where we have dropped the dependence on the control actions on the right hand side for notational clarity. When the number of estimated targets is less that 1 i.e. $\hat{n}^j_k <1$ the tracking score attains its maximum value. Finally we define the multi-agent tracking objective function to be minimized as $\sum_{j \in \alpha} \xi^j_\text{track}(u^j_{k},Z^j_{k|k-1,u^j_{k}})$ where $\alpha \subseteq S$ denotes the set of agents in \textit{tracking} mode. 

\subsection{Multi-Agent Searching Objective}
In the previous sub-section we have defined the objective function for the task of tracking for single and multiple agents. In this section we are going to do the same for the task of searching. Let us define the \textit{search value} at time $k$ for specific location $\textit{p} = (\textit{p}_x,\textit{p}_y) \in \mathcal{A}$ for the agent $j$ just after control action $u^j_{k} \in \mathbb{U}^j_{k}$ has been applied as follows:
\begin{equation}
\xi^j(\textit{p}, u^j_{k}) = 1-p_D(\tilde{\textit{p}},u^j_{k})
\end{equation}
where the function $p_D(\tilde{\textit{p}},u^j_k)$ is the agent's sensing model and $\tilde{\textit{p}} = [\textit{p}_x, 0, \textit{p}_y,0]$ constructs a location vector compatible with Eqn. (\ref{eq:sensing_model}). The idea of defining the search value as the complement of the probability of detection reflects the fact that locations with low probability of detection, should have high value for searching. As a consequence distant locations with respect to the agent's location appear to have increased search value i.e. these areas are worth exploring in order to find new targets. Assuming pairwise detection independence between all agents and all targets, the search value at location $\textit{p}$ when accounting for a set of agents $\alpha \subseteq S$ is given by: 
\begin{equation}\label{eq:searchValueField}
\xi^\alpha(\textit{p},\{u^j_k : j \in \alpha\}) = \underset{j \in \alpha}{\prod}~ \xi^j(\textit{p}, u^j_{k})
\end{equation}
where $u^j_{k} \in \mathbb{U}^j_{k}$. The total search value over the surveillance area, i.e. the multi-agent search objective function is now given by:
\begin{equation}\label{eq:sv_total}
\xi^\alpha_\text{search}(\{u^j_k : j \in \alpha\})=\frac{1}{\textit{A}}\int_\mathcal{A} \xi^\alpha(\textit{p},\{u^j_k : j \in \alpha\}) d\textit{p}
\end{equation} 

\noindent where $\mathcal{A} \subset\mathbb{R}^2$  and \textit{A} is the total area of the 2D surveillance region. 

\begin{figure}
	\centering
	\includegraphics[width=\columnwidth]{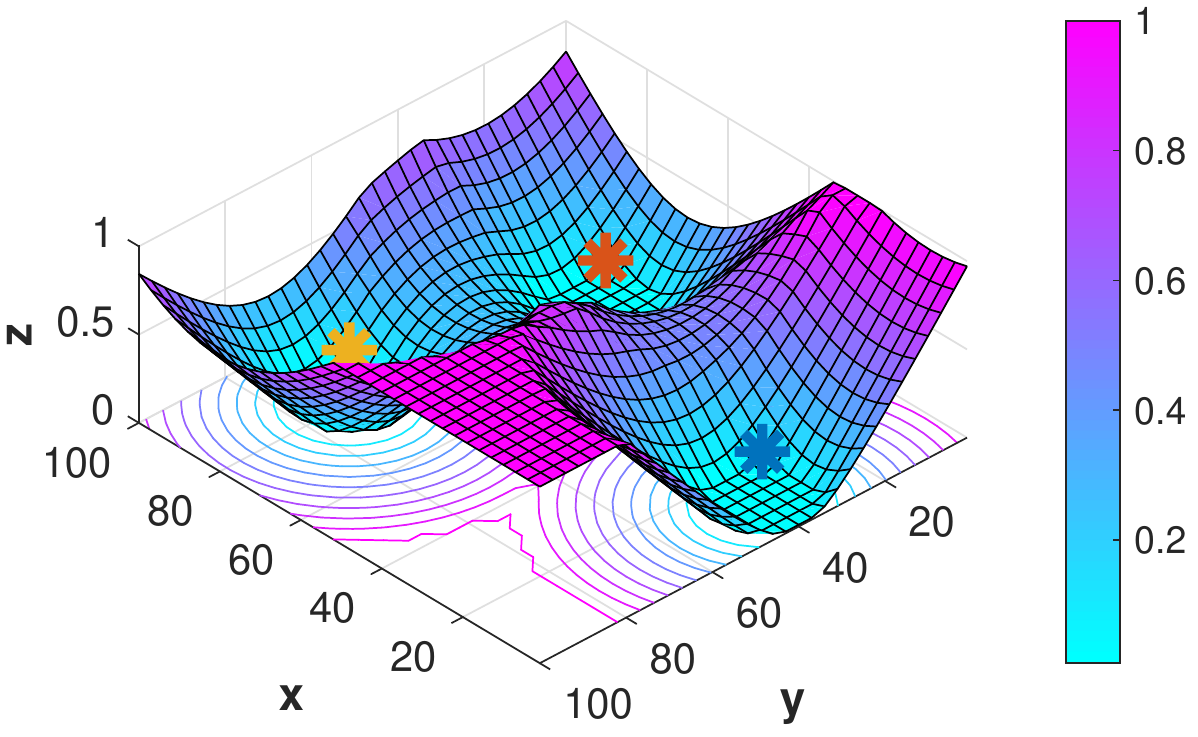}
	\caption{The figure shows an illustrative example of the search value i.e. Eqn. (\ref{eq:searchValueField}) over a surveillance area of size 100m by 100m for 3 agents. The position of agents are marked with $\ast$.}
	\label{fig:sv}
\end{figure}

\subsection{Implementation}
The optimization problem $(P_1)$ in Eqn. (\ref{eq:o1}) requires the enumeration of all possible combinations of the the power set $\mathbb{P}(S)$ and joint control actions between agents, in order to select the optimal modes of operation and controls which result in the overall minimization of the search-and-track objective.
 Clearly, this search space grows exponentially with the number of agents and the number of controls, resulting in a very hard combinatorial problem. Instead of solving this problem directly, we use the intuition behind $(P_1)$ to derive a heuristic algorithm that can solve instances of the problem in real time. To achieve this we approximate the mode selection as follows: At each time-step $k$ an agent $j$ can switch to tracking mode when $\sum_i r^{ji}_{k}(u,Z) \ge 1$ i.e. an agent switches to tracking mode only when the expected number of targets inside its sensing range is at least 1. Otherwise, the agent is in searching mode. In essence, this approximation helps us reduce the search space by taking into account information regarding the existence of targets in the area. The agents start by default in search mode and take joint mobility control decisions which minimize the search objective expressed in Eqn. \eqref{eq:sv_total}. Once an agent detects the presence of targets it switches to tracking mode and the control actions are due to the tracking objective function. An agent $j$ can switch back to search mode when $\sum_i r^{ji}_{k}(u,Z) < 1$. With the above heuristics however the system cannot attain the optimal behavior provided by the joint objective expressed in Eqn. \eqref{eq:o1}. For instance, an agent in tracking mode does not have the potential to switch to search mode while tracking targets. However, this approximation enables simplified switching that can be followed logically by the agents and which performs reasonably well in many scenarios.

\section{Evaluation}
\label{sec:Evaluation}

\subsection{Simulation Setup}
In order to evaluate the performance of the proposed approach we have conducted several numerical experiments involving various number of agents and targets for a variety of scenarios. In each experiment we compare the proposed approach with the ground-truth either qualitatively or quantitatively. In this section we include a representative sample of the conducted numerical simulations. 

More specifically, we assume that the targets maneuver in an area of 100m $\times$ 100m and that the single target state at time $k$ is described by $x_k = [\text{x},\dot{\text{x}},\text{y},\dot{\text{y}}]^\top$ i.e. position and velocity components in $xy$-direction. For the target dynamics i.e. Eqn. (\ref{eq:single_dynamics}) we assume that the target motion is piecewise linear according to the near constant velocity model with the process noise being Gaussian. Thus the single target transitional density is given by $\pi(x_k|x_{k-1}) = \mathcal{N}(x_k;Fx_{k-1},Q)$ where:
\begin{equation*}
	F = \begin{bmatrix} 
        1 & T & 0 & 0\\
        0 & 1 & 0 & 0\\
        0 & 0 & 1 & T\\
        0 & 0 & 0 & 1\\
       \end{bmatrix}, ~
    Q = \begin{bmatrix} 
        T/3 & T/2 & 0 & 0\\
        T/2 & T & 0 & 0\\
        0 & 0 & T/3 & T/2\\
        0 & 0 & T/2 & T\\
       \end{bmatrix}
\end{equation*}
with sampling interval $T=1$s. The target survival probability from time $k-1$ to time $k$ is constant $p_{s,k}(x_{k-1})=0.99$ and does not depend on the target's state. Once an agent detects a target it receives bearing and range measurements thus the measurement model of  Eqn. (\ref{eq:meas_model}) is given by:
\begin{equation*}
	h_k(x_k,s_k) = \left[ \text{arctan}\left(\frac{s_y-\text{y}}{s_x - \text{x}}\right),~ \norm{s_k -\text{H}x_k}_2\right]
\end{equation*}
where $\text{H} = \begin{bmatrix} 1 &0 &0 &0\\0 &0 &1 &0 \end{bmatrix}$. The single target likelihood function is then given by $g(z_k|x_k,s_k) = \mathcal{N}(z_k;h_k(x_k,s_k),\Sigma^\top \Sigma)$ and $\Sigma$ is defined as $\Sigma = \text{diag}(\sigma_\phi,\sigma_\zeta)$. The standard deviations $(\sigma_\phi,\sigma_\zeta)$ are range dependent and given by:
\begin{align*}
    \sigma_\phi &= \phi_0 + \beta_\phi \norm{s_k-\text{H}x_k}_2 \\
	\sigma_\zeta &= \zeta_0 + \beta_\zeta \norm{s_k-\text{H}x_k}_2^2
\end{align*}
with $\phi_0 = \pi/180$rad, $\beta_\phi=10^{-5}\text{rad}/\text{m}$, $\zeta_0 = 1$m and $\beta_\zeta = 3\times10^{-3}\text{m}^{-1}$. 
Moreover the agent receives spurious measurements (i.e. clutter) with fixed Poisson rate $\lambda_k = 5$ uniformly distributed over the measurement space. 
The agent's sensing model parameters take the following values $p_D^{\text{max}} = 0.99$, $\eta = 0.003$ and $R_0=10$m.
The agent's dynamical model has radial displacement $\Delta_R=2$m, $N_R=2$ and $N_\theta = 8$ which gives a total of 17 control actions, including the initial position of the agent. 
Finally, in order to handle the non-linear measurement model we have implemented a Sequential Monte Carlo (SMC) version \cite{Vo2009} of the multi-Bernoulli filter.

\subsection{Results}
\begin{figure*}
	\centering
	\includegraphics[width=\textwidth]{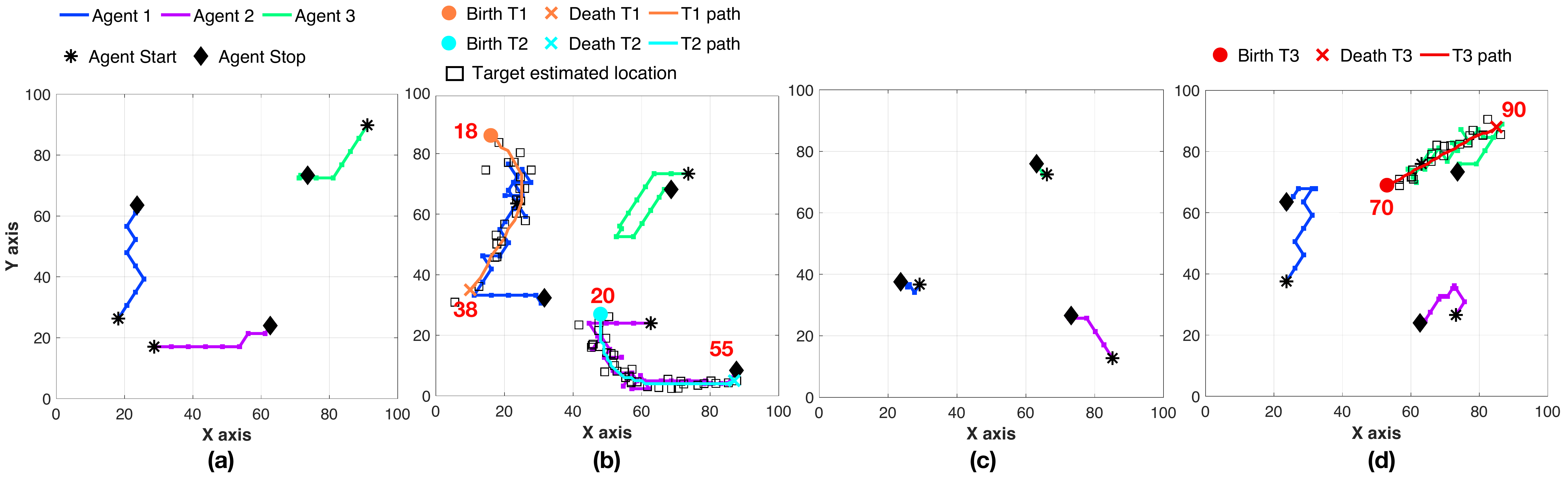}
	%\vspace{-7mm}
	\caption{The figure shows the maneuvers of 3 agents for the task of search-and-track in a simulated scenario during 4 time periods i.e. (a) - (d). Agent start and stop positions are marked with $\star$ and $\Diamond$, respectively. Target birth and death positions are marked with $\bullet$ and $\times$ respectively and their corresponding birth/death times are shown in red.}
	\label{fig:eval_sys1}
\end{figure*}

\begin{figure}
	\centering
	\includegraphics[width=\columnwidth]{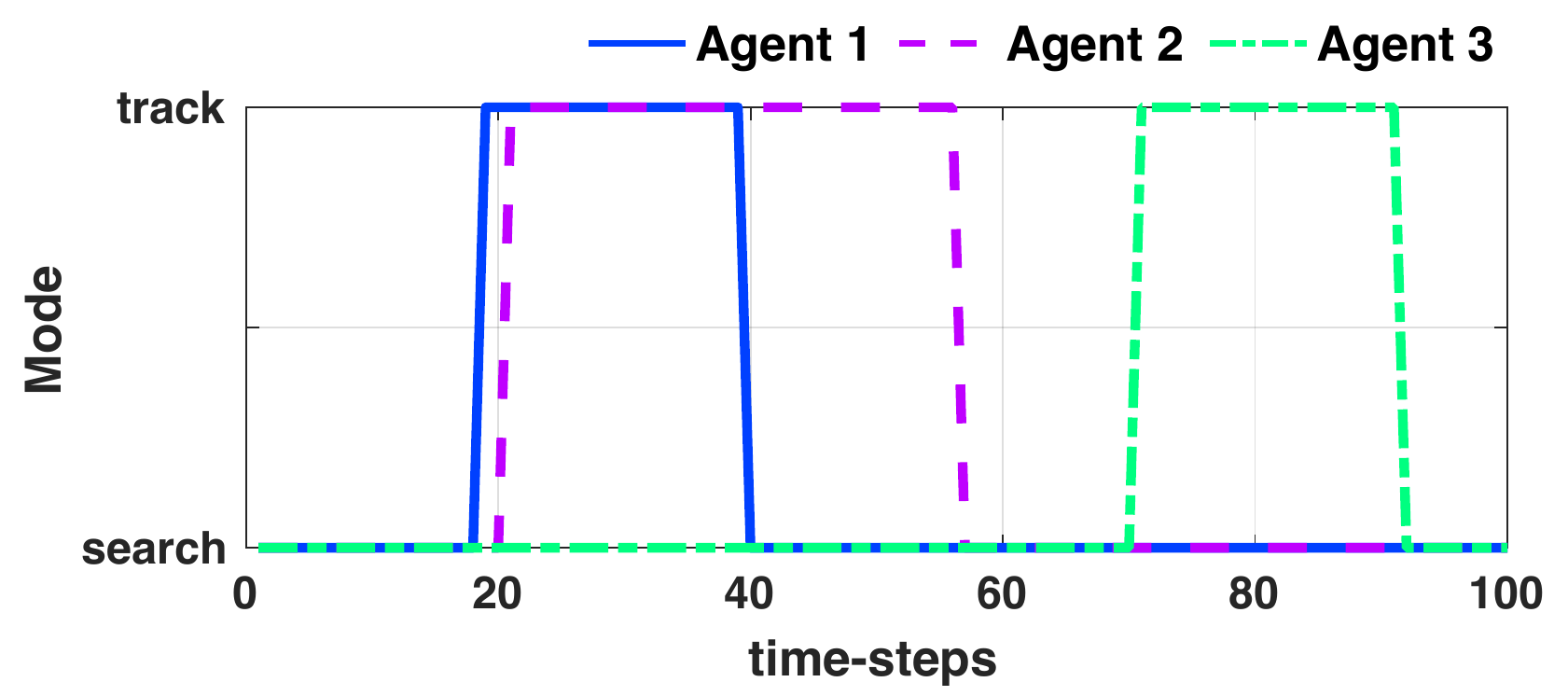}
	\caption{The figure shows the mode of operation i.e. search or track for the 3 agents of the simulated scenario shown in Fig. \ref{fig:eval_sys1}}
	\label{fig:modes}
\end{figure}

\textbf{Representative search and track scenario:}
We begin the evaluation of the proposed technique with a representative search-and-track scenario with 3 agents and 3 targets which takes place during 100 time-steps. This scenario is depicted in Fig. \ref{fig:eval_sys1} during 4 consecutive time periods i.e. $k=1..17$, $k=18..56$, $k=57..69$ and $k=70..100$. In this scenario, 3 agents enter at $k=1$ the surveillance area of size $100\text{m} \times 100\text{m}$ at the locations marked with $\star$ in Fig. \ref{fig:eval_sys1}a with coordinates $[18, 26]$, $[28, 17]$ and $[90, 89]$ for agents 1, 2 and 3, respectively. The target birth/death times are $k=18/38$, $k=20/55$ and $k=70/90$ for targets 1, 2 and 3, respectively. The target birth locations (marked with $\bullet$) are $[16, 86]$, $[48, 27]$ and $[53, 69]$ for targets 1, 2 and 3 respectively and their corresponding death locations (marked with $\times$) are $[10, 35]$, $[87, 5]$ and $[85, 88]$ as shown in Fig. \ref{fig:eval_sys1}. The trajectories of the agents and the targets during this experiment are shown in Fig. \ref{fig:eval_sys1}a -  Fig. \ref{fig:eval_sys1}d for periods $k=1..17$, $k=18..56$, $k=57..69$ and $k=70..100$ respectively. 
At each time-step the proposed system partitions the agents into the two modes (i.e. \textit{search} or \textit{track}) and selects the control actions for all agents that optimize the overall search-and-track objective as explained in the previous sections.

Figure \ref{fig:eval_sys1}a shows the trajectories of the 3 agents during period $k=1..17$. During this period the number of estimated targets is 0 so the system assigns all agents to search mode for the whole period. The objective now is to increase coverage by optimizing the search objective i.e. Eqn. \ref{eq:searchValueField}. As we can observe from the figure agent 1 and agent 2 start at $k=1$ to move away from each other in order to decrease the overlap in their sensing ranges. In particular, the locations of the 3 agents at the end of this period are $[23, 63]$, $[62, 23]$ and $[73, 73]$ respectively. This formation (i.e. agent positions) increases the area coverage since it reduces the sensing range overlap. The modes of operation for the 3 agents for this scenario is shown more clearly in Fig. \ref{fig:modes}.

Now, at time $k=18$ target 1 appears in the area as shown in Fig. \ref{fig:eval_sys1}b which is being detected by agent 1 at the next time-step $k=19$. During time-steps $k=19..23$ agent 1 switches to tracking mode and moves towards target 1 to intercept as shown in the figure. Between time-steps $k=24..38$ agent 1 tracks target 1. At the same time target 2 appears in the surveillance region at time-step $k=20$. This target is subsequently detected and tracked by agent 2. Now, during time-steps $k=21..38$ agents 1 and 2 are in track mode which leaves only agent 3 in search mode. For this reason agent 3 moves toward the center of the surveillance area to increase coverage. However, at time-step $k=39$ target 1 dies and thus agent 1 switches to search mode as well. At this point agent 1 and agent 3 are in search mode while agent 2 is in track mode. As we can observe from the figure at $k=39$ agent 3 moves away from the center of the surveillance region in order to jointly optimize coverage along with agent 1. Meanwhile agent 2 tracks target 2 until $k=55$ as shown in the figure.

Next, at $k=56$ target 2 dies, which makes agent 2 to switch to search mode. Thus during the 3rd period i.e. $k=57..69$ all agents are in search mode. From Fig. \ref{fig:eval_sys1}c we can see that agents 1 and 3 make slight adjustments in their positions whereas agent 2 moves away from the edge of the surveillance region in order to jointly optimize with the other agents the area coverage. 

Finally, during the 4th period at $k=70$ target 3 appears which is subsequently detected and tracked by agent 3 during time-steps $k=71..90$ as is shown in Fig. \ref{fig:eval_sys1}d. Meanwhile, agents 1 and 2 remain in search mode and move towards the diagonal of the square surveillance area which results in maximum coverage. To conclude the experiment, at $k=91$ agent 3 switched back to search mode after target 3 dies and all agents optimize the area coverage as is shown in the figure.

\begin{figure}
	\centering
	\includegraphics[width=\columnwidth]{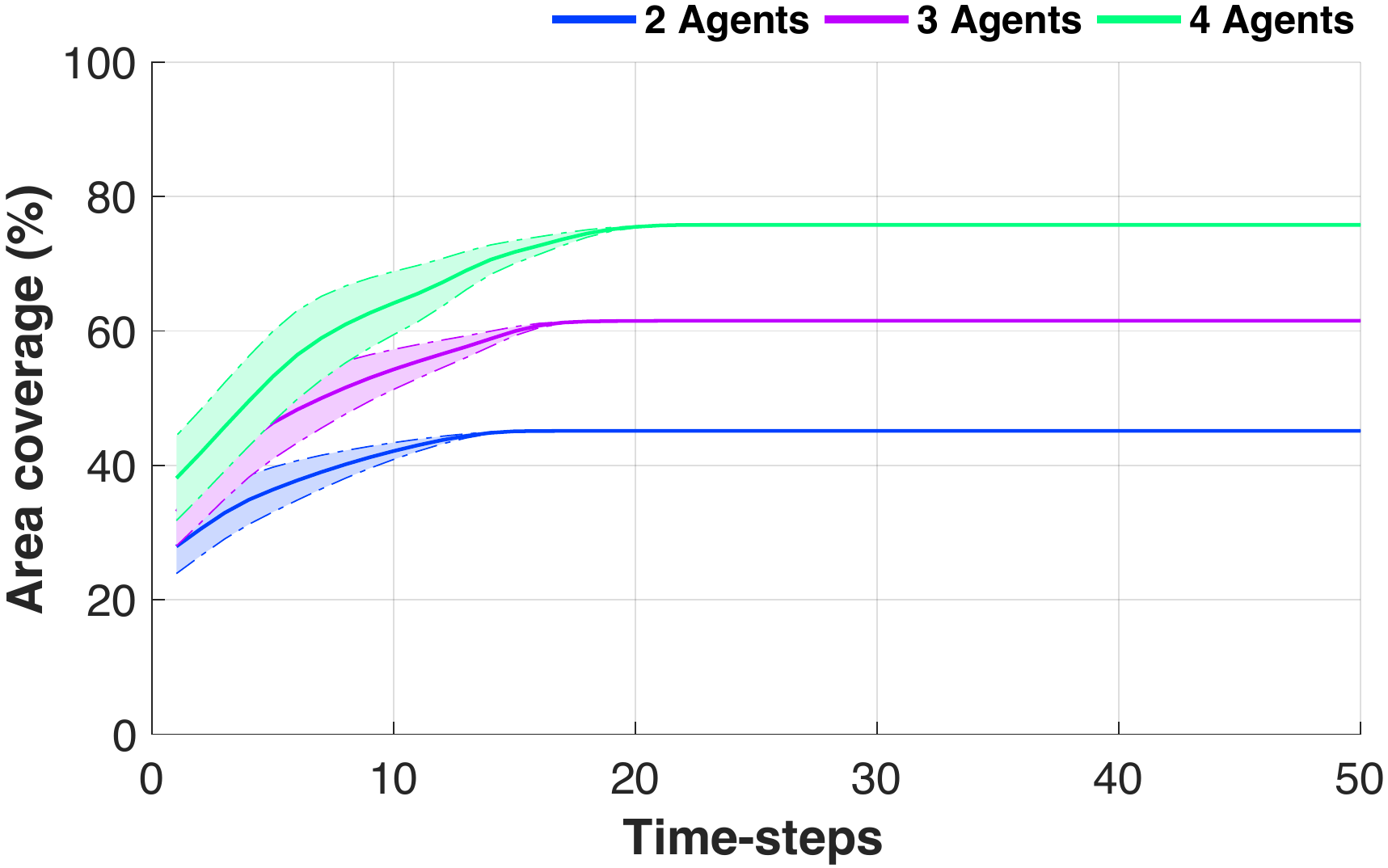}
	\caption{The figure shows the percentage of covered surveillance area over time for different number of agents.}
	\label{fig:fig2}
\end{figure}

\textbf{Multi-agent searching:}
The next set of experiments aims to investigate in more detail the behavior of our system when all agents are in search mode. More specifically, the desired behavior is for the agents to jointly find the best configuration i.e. select the appropriate controls such that the total search value of the surveillance area is minimized or the area coverage is maximized. To investigate this we have performed the following procedure. For a given number of agents we have conducted 50 Monte-Carlo (MC) trials where the agents are spawned from random locations (uniformly distributed) inside the surveillance area of size 100m $\times$ 100m. Then we let our system to run for 50 time steps and we monitor the percentage of covered area. Our intuition is that the search value should decrease over time since the agents would try to move to locations where their sensing range overlap is minimized which in turn will result in an increase in the area coverage over time. In other words we would like to have the agents spread as much as possible inside the surveillance region to increase coverage. Figure \ref{fig:fig2} shows the average area coverage over time for 2, 3 and 4 agents (with $p_D^{\text{max}} = 0.99$, $\eta = 0.03$ and $R_0=10$m). As we can observe from this figure the search value is decreasing over time which is evident by the increase of the covered area over time. The agents are trying to cover as much area as possible and minimize the overlap in their sensing ranges. It is also clear that as the number of deployed agents inside the surveillance area increases the covered area increases as well.

\begin{figure}
	\centering
	\includegraphics[width=\columnwidth]{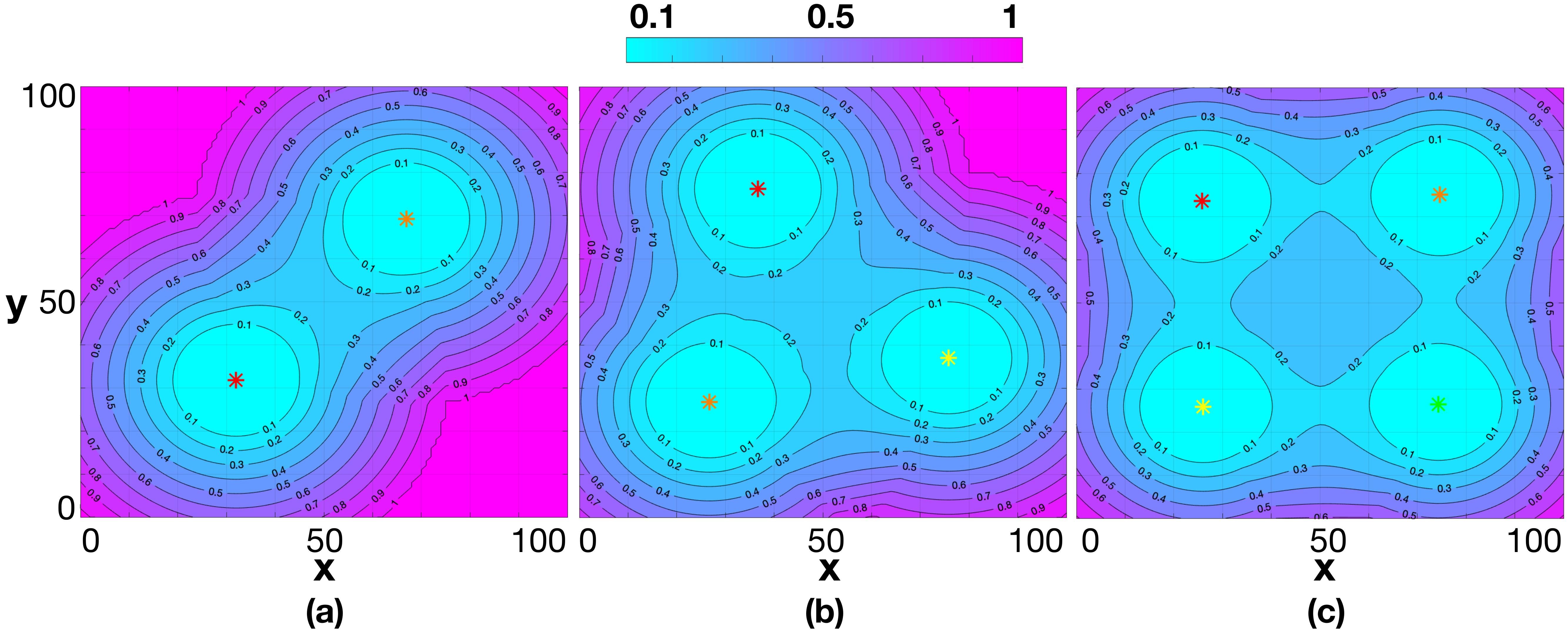}
	\caption{The figure shows the search value over the whole surveillance area for different number of agents, i.e (a) 2 agents, (b) 3 agents and (c) 4 agents. As we can observe the agents take positions which result in the maximum area coverage (the figure shows the last time-step of a 50 time-step simulation experiment).}
	\label{fig:fig4}
\end{figure}

In order to investigate further the experiment conducted above we have being monitoring the controls assigned to the agents during the search. At the end of the experiment the final agent positions for the 3 configurations i.e. 2, 3 and 4 agents are shown in Fig. \ref{fig:fig4}. Interestingly, we can see that the agents have converged to a formation which results in maximum coverage. For instance from Fig. \ref{fig:fig4}a we see that when the number of agents is 2 the configuration which results in the maximum coverage is the diagonal of the square surveillance area. On the other had when we have 3 and 4 agents, the optimal configuration forms a triangle (Fig. \ref{fig:fig4}b) and a square respectively (Fig. \ref{fig:fig4}c). This behavior however depends on the agent's sensing model (i.e. Eqn. (\ref{eq:sensing_model})). It is not unlikely to have other formations given a different sensing model. This is something quite interesting that we will investigate in the future. In addition, it would be very interesting to see that happens when not all agents possess the same sensing model. This is also left for future work. 

We should note here that in the previous experiment no targets were initiated during the simulation. This was done in order to investigate the behavior of the system in the search mode. Additionally, the number of targets and their initial locations are not known a priori in this work, instead there are being estimated at run-time. Thus during the previous experiment, the proposed system correctly assigned all agents in search mode since no targets where present.

\begin{figure}
	\centering
	\includegraphics[width=\columnwidth]{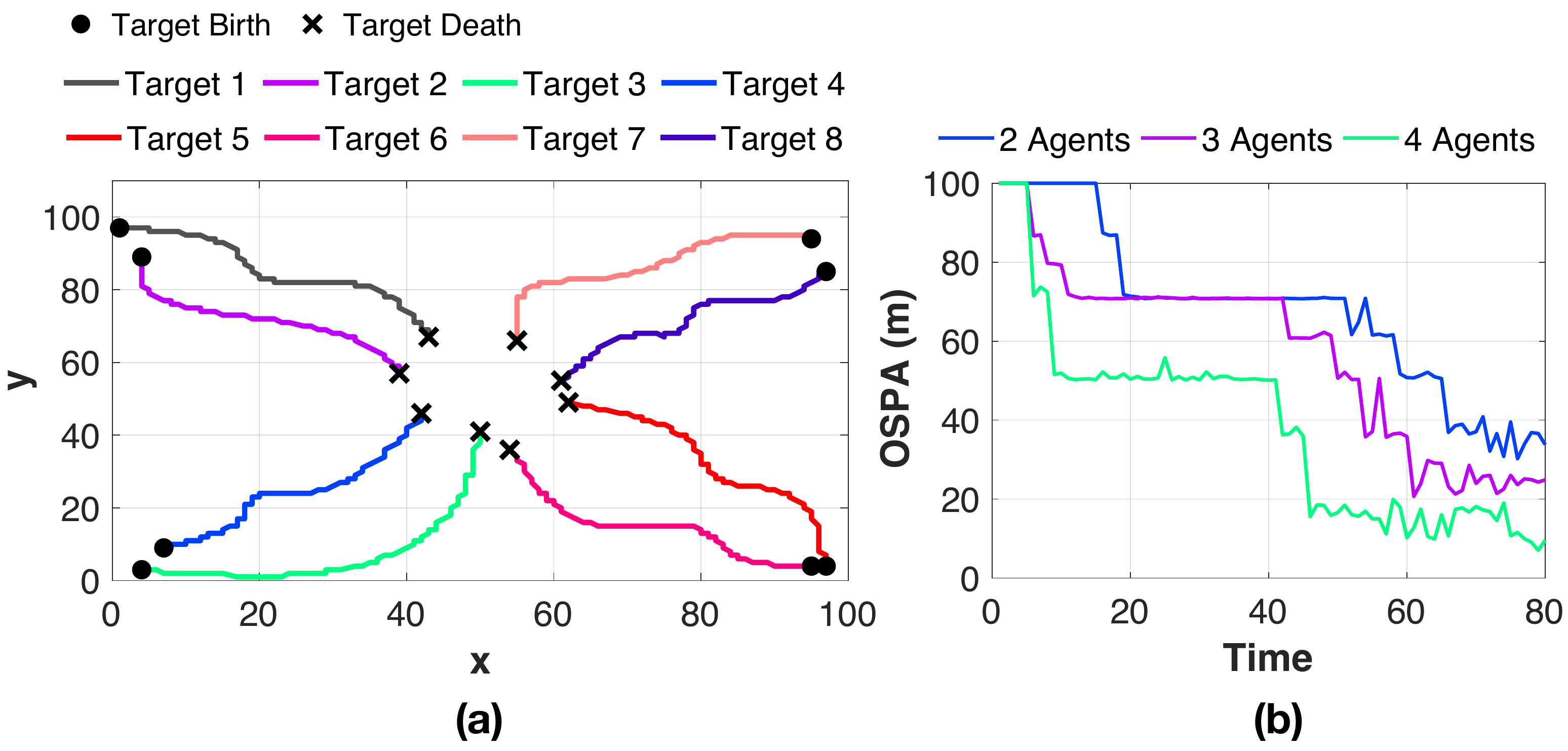}
	\caption{The figure demonstrates the search-and-track performance of the proposed technique through the OSPA error.}
	\label{fig:fig1}
\end{figure}

\textbf{Search and track performance:}
Finally, our last experiment was conducted in order to demonstrate the search-and-track performance of the proposed system. For this experiment we have used the optimal sub-pattern assignment (OSPA) metric \cite{Schuhmacher2008} which is widely used to measure the accuracy of multi-object filters as an indicator of the search-and-track performance. In this experiment which takes place during 80 time-steps, 8 targets are spawned at the following locations $[1, 97]$, $[4, 89]$, $[3, 4]$, $[7, 9]$, $[97, 4]$, $[95, 4]$, $[95, 94]$ and $[97, 85]$ for targets 1 to 8 respectively and die at $[43, 67]$, $[39, 57]$, $[50, 41]$, $[42, 46]$, $[62, 49]$, $[54, 36]$, $[55, 66]$ and $[61, 55]$ as shown in Fig. \ref{fig:fig1}a. The target birth/death times are $1/80$ for all targets. With this setup 2, 3 and 4 agents are spawn simultaneously at time-step $k=1$ from the center of the surveillance area (i.e. the agents are spawned randomly inside the bounding box $[40..60]\times[40..60]$) and the system runs for 80 time-steps. We perform 30 MC trials for each configuration (i.e. 2, 3 and 4 agents) and we measure the OSPA error. Figure \ref{fig:fig1}b shows the average OSPA error (with parameters $c=100, p=2$) for the 3 configurations. 

The first observation to make is with respect to the searching behavior of our system. Since all agents start from about the center of the surveillance area whereas all targets start from the perimeter there is a period of searching during the first time-steps of the experiment. Indeed, during the first time-steps all agents are in search mode and try to maximize the area coverage. This is shown in Fig. \ref{fig:fig1}b from the OSPA error which takes the maximum value during the beginning of the test for all agent configurations. When the number of targets being tracked is 0 the OSPA error reaches its maximum value which is 100m in this case. Moreover, we observe that with 3 and 4 agents the search is faster compared to the 2 agent case. With 2 agents the first target(s) seems to be detected at around $k=15$, whereas 3 and 4 agents detect the first targets around $k=5$. This is reasonable and verifies our intuition that as the number of agents increases the search performance should increase as well. 
The second observation from this experiment is with respect to the tracking capacity of the proposed system. As we can observe from the OSPA error as the number of agents increases the tracking accuracy increases as well. Since each agent is capable of tracking multiple targets the overall tracking accuracy can be increased by positioning the agents in locations which will result in maximum target coverage. This is achieved in this work by minimizing the sensing overlap between agents as is shown in this experiment.

\section{Conclusion and Future Work}

\label{sec:Conclusion}
In this paper we have studied the problem of searching and tracking of an unknown and time-varying number of targets with a team of mobile agents. We have presented a novel probabilistic framework based on random finite sets that accounts for the uncertainty in the number of targets, target dynamics and measurement models. The proposed approach allows the agents to optimize the search-and-track objective by dynamically switching their mode of operation at each time-step during the mission according to the acquired information. Finally, we have demonstrated the performance of the proposed decision and control algorithm through extensive numerical simulations. Future work will investigate the extension of this approach in 3D environments and study the problem of searching and tracking in highly dynamic environments with obstacles and occlusions.  

% Finally, in this work we assumed that targets and agents exist in a 2D world (e.g. UAVs with fixed and not controllable altitude). The extension of this work in 3D environments will be investigated in future works.
%Future work will look at simpler and computationally improved approximations of the non-linear integer program as well as alternative and simpler objective functions with equivalent behavior and performance. In addition, we are interested in investigating a rolling horizon (i.e. multiple-step look ahead) predictive control approach for this problem and analyzing its performance against the proposed single-step look ahead system.

\section*{Acknowledgments}
\noindent This work is supported by the European Union Civil Protection under grant agreement  No 783299  (SWIFTERS), by the European Union’s Horizon 2020 research and innovation programme under grant agreement No 739551 (KIOS CoE) and from the Republic of Cyprus through the Directorate General for European Programmes, Coordination and Development.
%\balance
%\raggedbottom
\flushbottom
\balance
\bibliographystyle{IEEEtran}
\bibliography{IEEEabrv,icuas19}  
%\raggedbottom

\end{document}